# Near–field magnetostatics and Neel–Brownian interactions mediated magnetorheological characteristics of highly stable nano–ferrocolloids

Ajay Katiyar[1, 2,*], Purbarun Dhar[2, #], Sarit K. Das[2, $] and Tandra Nandi[3]

[1] Research and Innovation Centre (DRDO), Indian Institute of Technology Madras Research Park,
Chennai–600 113, India

[2] Department of Mechanical Engineering, Indian Institute of Technology Madras,
Chennai–600 036, India

[3] Defense Materials and Stores Research and Development Establishment (DRDO), G.T. Road,
Kanpur–208 013, India

[#]*Electronic mail:* pdhar1990@gmail.com
*Corresponding author: Electronic mail:* ajay_cim@rediffmail.com
*Tel No: 044-22548-222*
*Fax: 044-22548-215*
[$]*Corresponding author: Electronic mail:* skdas@iitm.ac.in
[$]*Phone: +91-44-2257 4655*
[$]*Fax: +91-44-2257 4650*

# Abstract

Magnetic nanocolloids comprising of synthesized super paramagnetic $Fe_3O_4$ nanoparticles (SPION) (5–15 nm) dispersed in in–situ developed Polyethylene Glycol (PEG-400) and nano-silica complex have been synthesized. The PEG–nano Silica complex physically encapsulates the SPIONs, ensuring no phase separation under high magnetic fields (~1.2 T). Exhaustive magneto-rheological investigations have been performed to comprehend the structural behavior and response of the ferrocolloids. Remarkable stability and reversibility have been observed under magnetic field for concentrated systems. The results exhibit the impact of particle concentration, size and encapsulation efficiency on parameters such as shear viscosity, yield stress, viscoelastic moduli, magnetoviscous hysteresis etc. Analytical models to reveal the system mechanism and mathematically predict the magneto–viscosity and magneto–yield stress has been theorized. The mechanistic approach based on near-field magnetostatics and Neel-Brownian interactivities can predict the colloidal properties under the effect of field accurately. The colloid exhibits amplified storage and loss moduli alongside highly augmented linear viscoelastic region under magnetic stimuli. The transition of the colloidal state from fluidic phase to soft condensed phase and its viscoelastic stimuli under the influence of magnetic field has been explained based on the mathematical analysis. The remarkable stability, magnetic properties and accurate physical models reveal good promise for the colloids in transient situations viz. magneto–MEMS/ NEMS devices, anti–seismic damping, biomedical invasive treatments etc.

**Keywords:** Magnetic nanofluids; ferrocolloids; super paramagnetism; $Fe_3O_4$ nanoparticles; viscosity; magnetorheology; yield shear stress; reversibility; soft magnetocolloids

# 1. Introduction

'Smart fluids' has been an area of academic interest, research and development that has enjoyed widespread reception from the academic community over the past two decades. A plethora of active research endeavors are aimed towards discovering 'smarter' fluids and towards understanding their underlying physics. Such fluids involve Magnetorheological (MR) fluids[1, 2], Electro rheological (ER) fluids[1], Shear Thickening fluids, thermo–nanofluids, optically smart fluids and ferrofluids to name a few categories. Among these, ER and MR fluids have received a greater part of the attention owing to their earlier discoveries. Among the two, the MR fluids have been more successful and have also been commercialized. MR fluids[3], as the name suggests, are special fluids whose viscous and rheological properties respond to externally applied magnetic stimulus. Such fluids involve micron or mesoscale ferromagnetic or ferromagnetic particles[4, 5] dispersed in a base carrier fluid, often aided by a stabilizer so as to prevent phase separation during operation or storage. They differ from ferrofluids[6, 7, 8] with respect to particle size and concentration, the latter consisting of nanoscale magnetic particles with low particle concentration. Although ferrofluids lack the ability to sustain loads even at high magnetic stimuli, owing to their dilute nature, they enjoy very long stability and non–erosive operation due to the nanoscale particles which are in random motion due to thermal fluctuations [9]. Thereby, keeping in mind the trend of miniaturization and longer operating life, it becomes a necessity to develop a new class of ferrocolloids involving nanoscale particles such that both the advantageous aspects of MR as well as ferrofluids can be exploited with a single fluid. The present research article discusses the development of such a super paramagnetic iron oxide nanoparticle (SPION) based ferrocolloid with very high stability and magnetic response, the involved physical properties and detailed mathematical analysis of the underlying mechanisms involved.

An inherent problem with conventional MR fluids is settling of the constituent particulate phase with time. The present system uses nanoscale magnetic particles instead of conventional micron scale particles, thereby increasing the stability. Furthermore, the PEG–nSi complex forms a singular entity with infinite shelf life. The magnetic particles trapped within this system thereby yield ferrocolloids with shelf lives exceeding years. Conventional MR fluids also suffer the problem of erosion during action due to the presence of micron size particles. The presence of nanoscale particles in the present case therefore overcomes this problem altogether. The method of chemical or physical encapsulation to prevent the magnetic particles from separating out of the base matrix at high field strengths has been used in a few known cases[10, 11], but has been restricted to mesoscale particles. The procedure involves separately coating the particles with a protective layer before dispersing them in the carrier fluid. However, if a methodology wherein the carrier matrix could itself act as a coating agent could be developed; the synthesis procedure would be less time consuming, feasible and could be replicated in bulk, since the intermediate step of coating the particles is absent. The present study involves a novel methodology wherein the base matrix itself acts as the protective coating agent and provides long term stability to the system, without compromising its magnetic properties. Furthermore, the magnetoviscous properties observed in the present class of colloid is superior in magnitude and applicability than similar colloids which involve nanoscale iron[12] or nickel/cobalt[13] particles as the dispersed magnetic media[14]. Since production of size controlled SPIONs are relatively easier, less expensive and do not suffer the disadvantages of oxidation during use than metallic magnetic nanoparticles, the present colloids are superior from materials as well as utility point of view.

The present study also focuses on predicting the magnitudes of application oriented rheological and viscoelastic properties[15, 16] of a ferrocolloid based on a comprehensive understanding of the underlying magneto–hydrodynamic mechanisms at play, nanomaterial characteristics and intuitive analyses of the same. Based on an application standpoint, the properties of viscosity and yield stress as a function of SPION concentration and size and applied magnetic field are among the most important for design of ferrocolloids for specific targeted systems. The present manuscript also incorporates an intuitive analytical methodology to determine the effect of magnetic field strength on the propensity of chain formation in a ferrocolloid of known SPION concentration and its impact vis-à-vis enhanced viscosity and yield strength. A mathematical analysis on the behavior of the viscoelastic moduli with magnetic flux density and detailed discussions on the internal mechanisms responsible for the observations have been incorporated. In précis, the article reports a novel methodology to form highly stable SPION based ferrocolloids, its magnetoviscous and magnetoviscoelastic properties as a function of SPION concentration and magnetic field intensity and provides analytical models and detailed insight into the physics behind the observed behaviors.

## 2. Experimentation : Materials and Methodologies

### 2.1 Synthesis and characterization of Fe₃O₄ nanoparticles

In the present work, $Fe_3O_4$ nanoparticles have been prepared by a modified co-precipitation method [17]. Co-precipitation is a facile and relatively easy method for synthesis of iron oxide nanostructures ($Fe_3O_4$ or $\gamma$-$Fe_2O_3$) from aqueous ferrous and ferric salt solutions by reduction via a base. The source of magnetite particles in the present case are anhydrous Ferric Chloride (Analytical grade, Merck) and Ferrous Chloride tetrahydrate (Analytical grade, Sigma-Aldrich) and the reducing agent utilized is 30% Ammonia solution (Analytical Grade, Sisco Research Labs). The iron salts are separately dissolved in deionized water to obtain two 1 M stock salt solutions. The two solutions are then mixed in the volumetric ratio of $Fe^{3+}/Fe^{2+}$ = 2:1, stirred and heated to a temperature of ~ 75-85 °C. The ammonia solution is then slowly added to the hot salt solution while stirring with a mechanical stirrer at ~ 3500–4000 rpm. The solution turns dark brownish black and ammonia solution is added till the pH of the solution reaches 11-12. The solution is allowed to stand still overnight and then centrifuged to collect the precipitate. The precipitate is washed repeatedly with double distilled water till the pH of the solution turns neutral. The particles are then washed with Acetone (analytical grade, Merck) to remove any organic impurities and dried in inert atmosphere for 10–12 hours. The dried powder is then ground in a mechanical mortar and pestle under nitrogen atmosphere so as to reduce the agglomerations induced by the drying process. It has been observed that the additional heating of the iron salt mixture prior to reduction and the utilization of high speed stirring during the reaction leads to iron oxide nanoparticles of smaller size, higher sphericity and narrower size distribution when compared to the conventional co-precipitation methodology.

The synthesized nanoparticles have been characterized by Transmission Electron Microscopy (TEM) so as to determine their shape, size and structural morphologies. The TEM images of the particles and that of a single nanoparticle have been illustrated in Fig.1 (a) and (b) respectively. Analysis (Fig. 1(a)) reveals nanoparticles of uniform size and low polydispersity, with the size ranging from 5-20 nm and a mean size of ~ 10 nm. The particles also exhibit high degree of sphericity with very limited structural eccentricity (Fig. 1(b)). The size of the particles synthesized fall much below the critical single domain size of $Fe_3O_4$ particles (~ 128 nm)[17, 18] and therefore are super paramagnetic in nature. The crystalline structure of the super paramagnetic iron oxide nanoparticles (SPION) has been confirmed with Selected Area

Diffraction Pattern (SAED) characterization. (Inset Fig.1 (b)). Appearances of concentric diffraction rings prove the existence of pure crystal structure. The magnetic saturation value of the SPION is experimentally determined by a Vibrating Sample Magnetometer (VSM) and the M vs. H curve has been illustrated in Fig. 1(c). The particles attain a saturation magnetic moment of 51emu/g at 300 K at a saturation magnetic field of 12.5 KOe. The sample exhibits zero or negligible hysteresis which shows that the super paramagnetic behavior is preserved at low and reversible field strengths.

**FIGURE 1**

## 2.2. Preparation and characterization of magneto–nanocolloids

The ferrocolloids have been prepared by physically dispersing the synthesized SPION into an in-situ synthesized complex consisting of Polyethylene glycol (PEG-400, analytical grade, Sigma-Aldrich, dynamic viscosity 100-120 mPas), Silica nanopowder (spherical, average particle size 7 nm, specific surface area approx. 390±40 $m^2$/g, high purity, Sigma–Aldrich) and Ethanol (Absolute grade, Sisco Research Labs). All chemicals were utilized as received without further purification. Silica has been shown to fairly stabilize ferrofluids[19]; however, for concentrated ferrocolloids, a more stable system is obtained using the PEG–nSi complex. The preparation methodology involved two steps. The first step involves the preparation of the Polyethylene glycol-nano Silica (PEG-nSi) complex. The complex has been observed to play the most essential role in stabilizing the iron oxide nanoparticles in the colloidal state and prevents phase separation from the PEG even under the influence of strong magnetic fields. To begin with, nano Silica (0.5 wt. %) is dispersed in Ethanol (1:3) by ultrasonication for 30 minutes under favorable pulse amplitude and frequency. A Dynamic Light Scattering (DLS) analysis of the ethanol-silica solution diluted with water (Inset 3, Fig. 2) confirms the size of the nanostructures. The silica-ethanol solution is then added slowly to PEG (Si-Ethanol to PEG ratio = 1:3 v/v) under the influence of ultrasonication. The final fluid is allowed to be sonicated for 3-4 hours, until a semi-gel is formed. The PEG-nSi complex has been found to have infinite shelf life.

The second step involves dispersing the SPION in the PEG-nSi complex. The synthesized nanoparticles are first dispersed in Ethanol by sonication to form a stable colloid. Next, the complex is subjected to sonication and the ethanol-iron oxide colloid is slowly added to PEG-nSi while maintaining at room temperature. The final PEG-nSi-SPION colloid is sonicated for 3 hours to ensure complete encapsulation of the magnetic particles by the complex. The ferrocolloid is then heated to ~80 °C and allowed to stand at the same temperature for 3 hours so as to allow complete vaporization of the Ethanol (boiling point ~ 80 °C) from the colloid. The Ethanol acts as an intermediate solvent so as to reduce the viscosity of the complex and the colloid during preparation. A highly dilute version of the sample was prepared by dissolving a small amount of the ferrocolloid in ethanol for characterization purposes utilizing High Resolution Scanning Electron Microscopy (HRSEM). The HRSEM images; illustrated in Fig. 2, reveal the presence of a matrix of granulated PEG-nSi complex with the SPION dispersed within it. The SPIONs also exhibits a nanoscale coating of the granular complex, as distinguishable from the core SPION due to its lighter hue, as illustrated in Inset 2, Fig. 2. A schematic of the encapsulated system has been illustrated in Inset 4, Fig. 2. The PEG molecules form chained meshes on the surface of each silica nanoparticle, forming the PEG-nSi complex. The microstructure of the PEG-nSi complex is therefore a complex mesh of PEG molecules anchored from silica nanoparticles. During the ultrasonication process, it is proposed that the SPIONs get trapped within such meshed networks. This proposal is strengthened by the fact that the SPIONs do not separate from the colloid even under the influence of high magnetic fields. This is only possible if they have been physically trapped within the strong PEG-nSi complex network. Ferrocolloids of varying concentrations of SPION (5, 10, 12 and 15 wt. %) have been synthesized and extensive rheological and magneto-rheological and analytical studies have been performed to understand the response and behavior of the colloids.

**FIGURE 2**

## 2.3. Rheometry of the ferrocolloid

Steady and oscillatory rheological and magnetorheological experiments have been performed for the synthesized ferrocolloids utilizing a Physica MCR 301 rheometer (Anton Paar GmBH, Germany). The present experimentations involve parallel plate configuration with the plate separation maintained constant at 0.5 mm. All experiments have been performed at a constant temperature (298 K) maintained by a controlled temperature unit (Viscotherm VT2, Anton Paar, Germany). The rheometer was calibrated with the help of standard solutions to ensure accuracy and repeatability of the measurements. The maximum uncertainty involved was determined to be less than ~±2%. The rheometer consists of an additional magnetorheological controller to control the magnetic flux density via current control and a Teflon shield to reduce the magnetic flux losses during the experiments. The setup comprises electromagnetic coils placed exactly below the measuring plate such that it forms a toroid whose center coincides with the center of the measuring plate. The controller is capable of introducing current in the range of 0 to 5 A and the coil generates corresponding magnetic flux density ranging from 0 to 1.2 T. The controller and coil are calibrated using a standard gauss meter prior to the experimentations to obtain accurate magnetic flux density data.

## 3. Results and Discussions

### 3.1. Magneto-response of the ferrocolloid

The structural response of the ferrocolloids has been observed both visually as well as from rheometry data analysis for 0-15 wt. % concentration of SPION. Under normal circumstances, the ferrocolloid behaves like a viscous fluid (Fig. 3(a)) and exhibits very low rates of settling with the shelf life exceeding a year. However, in presence of a magnetic field, the super paramagnetic particles align along the direction of the field. It was observed that at 15 wt. % concentration of SPION, the magnetic field shows strong effect on the physical structure of the ferrocolloid (Fig. 3(b)) as compared to lower concentrations. Figures 3(c) and (d) provides a mechanistic illustration of the chain formation[20] and induced structural rigidity during transition from the cases illustrated in Fig. 3(a) to that of Fig. 3(b).

**FIGURE 3**

### 3.2. Field Induced Viscous Behavior of the Ferrocolloid

Experimental observations provide insight into the effects of SPION concentration and the external field strength on the viscosity of the ferrocolloids. Observations suggest that the magneto-viscosity is proportional to the SPION concentration as well as to the magnetic flux density. The increase in viscosity is caused by the formation of chained structures by the SPIONs in presence of the external field, which in turn provides increasing resistance to shear[16]. The maximum stable magneto-viscosity observed was at a SPION concentration of 15 wt. % at 1.0 T magnetic field (Fig. 4). It has also been observed that the ferrocolloids can efficiently sustain high magnetic fields of up to 1.2 T for concentrations up to 15 wt. % and no phase separation of SPIONs from the carrier fluid was observed in such cases. However, with increased concentrations, the magnetic responses of the colloid were too high and consequently lead to phase separation at magnetic field strengths well short of 1 T. Thereby, it can be argued that the ferrocolloids under present study have been optimally designed for stability and response. The experimental observations on the magneto-viscous effect of the colloid have illustrated in Fig. 4.

Although experimentally observed, the phenomenon of magneto-viscosity requires effective mathematical treatment so as to predict the viscosity of the ferrocolloid for any particular concentration, SPION size and properties and magnetic field strength. The present article theorizes an analytical approach to model the observed phenomena for predictive purposes and to reveal the physics and mechanisms at play. The initial step towards the analytical model requires a basic form of equation that can effectively relate the function parameter to the independent variable, in this case the magneto-viscosity and magnetic flux density respectively. Experimental observations[21] are suggestive of a quadratic or cubic polynomial function that relates the two parameters under discussion. In the present scenario, a base quadratic equation has been assumed as the initial step towards the mathematical model (the cubic form has not been considered in order to reduce complexities during the analysis process). The base form of the equation relating magneto-viscosity ($\mu_m$) and the magnetic flux density ($B$) can be expressed as[21]

$$\mu_m(B) = a_1 B^2 + a_2 B + a_3 \quad (1)$$

The next approach lies in deriving mathematical expressions for the coefficients of the equation from system properties such that the final equation is dimensionally consistent and can predict the magnitude of the magneto-viscosity. As the equation suggests, the coefficient '$a_3$' is in essence a parameter dimensionally equivalent to viscosity and is the ferrocolloid viscosity for zero magnetic flux density. Therefore, the coefficient is equivalent to the zero field viscosity of the ferrocolloid, '$\mu_{m,zf}$'. Eqn. (1) therefore takes a new form as

$$\mu_m(B) = a_1 B^2 + a_2 B + \mu_{m,zf} \quad (2)$$

The coefficients of the equation, namely, '$a_1$' and '$a_2$' need to be analytically and intuitively modeled such that the equation can predict the magnitude of the field induced augmented viscosity while being dimensionally consistent. Viscosity of a fluidic system is in itself a complex property that needs to be comprehended from the interactions at the molecular scale. In the present case, the enhancement in viscosity can be modeled analytically and intuitively from the mechanisms at play at the particle scale. One of the major factors governing the magneto-viscosity can be argued to be the magnetic field strength at which the dispersed SPIONs attain saturation magnetic moment. The saturation is an important property because it limits the enhancement in viscous behavior with increasing field. As the particulate phase reaches saturation, its ability to form structural chains of enhanced strength reduces, thereby limiting the maximal viscosity the colloid can attain. Dimensional analysis reveals that the coefficient '$a_2$' should possess the dimensions equivalent to product of the magnetic field strength and time and thereby can be modeled in accordance to the saturating magnetic field ($H_{sat}$) and the Neel Relaxation time ($\tau_N$) and a corresponding function of the SPION concentration. The dependence on $H_{sat}$ has been confirmed from present and published experiments[22] and the coefficient has been similarly deduced to be a function of the second power of SPION concentration. The dependence on the relaxation time has been explained in the following paragraph. The coefficient can be mathematically expressed as

$$a_2 = H_{sat} \tau_N \varphi^2 \qquad (3)$$

The coefficient '$a_1$' however requires a more complex formulation, as revealed from dimensional analysis, so as to infuse the field induced structure formation capability of the SPIONs within the mathematical model. Dimensional analysis shows that the coefficient is a function of some characteristic forcing function. This can be treated as the magnetizing force exerted upon individual particles which in turn lead to self-assembly and chain formation. The forcing function can be equivalently expressed as the ratio of a characteristic time scale and that of the magnetic permeability of the colloidal state. In case of SPION based magnetic systems, the most important time scale that can be considered is the Neel relaxation time scale[23]. In accordance to the single domain or macro-spin approximation, the SPIONs can change their direction of magnetization with the direction of the alien field. Due to the magnetic anisotropy, the magnetic moment in general possesses only two stable orientations, antiparallel to each other, separated by an energy barrier. The stable orientation defines the magnetic 'easy' axis of the SPION. At finite temperature, there exists a finite probability for the magnetization to flip and reverse its direction. Thereby, it is this flip time period that governs the ultimate strength of the chained structure formed under the influence of the external field. Smaller the flip period more is the probability for formation of stronger chains due to greater magnetic force of attraction between the individual SPIONs. The Néel relaxation time ($\tau_N$) is expressed by the Néel-Arrhenius equation[23, 24] as

$$\tau_N = \tau_0 \exp\left(\frac{KV}{k_B T}\right) \quad (4)$$

where, '$\tau_0$', '$k_B$', '$T$', '$K$' and '$V$' denotes the attempt period for the SPION material, the Boltzmann constant, the absolute temperature of the colloidal system, the magnetic anisotropy[25] of the SPIONs and the particle volume respectively. The magnitude of the attempt period and magnetic anisotropy for SPIONs has been adapted from published reports on the subject[26].

As discussed, the coefficient '$a_1$' can be mathematically expressed in terms of the Neel relaxation time, the magnetic permeability of the ferrocolloid, the average surface area and diameter of the SPIONs and the critical magnetic chain length under the influence of external field. The mathematical expression is as follows

$$a_1 = \left(\frac{\tau_N}{2\mu_0}\right)\left(\frac{A_s}{L_c d_p \varphi^n}\right)\left(\frac{d_{sep,\varphi}}{d_{sep,opt}}\right) \quad (5)$$

where '$A_s$', '$d_p$', '$L_c$', '$d_{sep,opt}$' and '$d_{sep,\varphi}$' represent the average surface area of the SPIONs, the average SPION diameter, the critical chained structure length under the influence of magnetic field, the interparticle separation within the chained structure at the optimal concentration and the separation at the required concentration respectively. The exponent '$n$' relates the magnitude of the coefficient to the concentration of SPION and It can be theorized that the total strength offered by the colloid in presence of magnetic field is inversely related to the critical length of the chains ($L_C$) formed. Formation of chains of long lengths implies reduced chain number density throughout the colloidal structure. It is noteworthy that augmented resistance to shear is introduced due to the formation of a homogenously distributed family of magnetic chain structures. Formation of long chains reduces the chain concentration while increasing the effective fluid phase, thereby leading to lower augmentation in viscosity. However, the preceding arguments are not sufficient to provide a mathematical backbone to predict the most probable chain size, which is a requirement to predict the augmented magneto-viscosity value. Therefore, there needs to be an allowable bounded value for the chain length, below which the chains cannot sustain themselves owing to Brownian fluctuations. The critical length so obtained can be utilized to deduce the magnitude of the coefficients.

The length of the chain structures can be deduced mathematically by applying the concept of thermally stable chain structures[27, 28]. Irrespective of the magnitude of the magnetic field, the particles in the present case (diameter ~ 10 nm) are prone to high degrees of thermal fluctuations due to their sizes[29]. Thereby, it deems necessary that the chains formed achieve such a length, such that they are, as a singular entity, thermally stable. This is an important requirement since thermally unstable chains cannot exhibit augmented shear resistance. From thermal stability considerations, the critical length of the field induced chain structures can be deduced from the Brownian diffusivity of the chain and their settling velocity as obtained from the Stokes-Einstein relation (from the densities of the suspended and the solute phase; $\rho_p$ and $\rho_f$, respectively) . The mathematical expression for the critical chain length is expressed as

$$L_c \geq \left( \frac{12 k_B T}{\pi (\rho_p - \rho_f) g} \right)^{1/4} \quad (6)$$

The present approach to deduce the most probable chain length based on intuitive analysis has been found to be accurate within ±15 % of the results obtained through extensive numerical computations[30].

The other important parameter that governs the magnetic moment of the chain structures as a whole is the mean inter-particle separation ($d_{c,sep}$) in the chain. This parameter is also expected to exhibit inverse variance with the viscosity. High separation values imply low fields and/or low particle concentration, either of which will lead to lower resistance to shear compared to high field and/or concentrations. Furthermore, low gaps are indicative of high magnetic polarization of the particles within the chain and thereby are less susceptible to thermal fluctuations within the chain. This leads to formation of stable chains which imply highest augmentations in shear resistance. Since modeling such a system involves statistical formulations, a very basic yet accurate approach involving linear assumptions can be resorted to. Since experimental observations can with ease help in identifying the optimal concentration for a stable and high response ferrocolloid, it can be assumed that it is at this concentration that the particles are in direct contact with each other while in the chained formation[31]. This assumption is justified since addition of further particles leads to decrease in shear resistance which evolves from particle over-crowding and phase separation. From knowledge of the optimal concentration it is thus possible to deduce the average inter-particle separation for lower concentrations from a simplified linear model expressible as

$$\frac{d_{sep,\varphi}}{d_{sep,opt}} = \frac{\varphi_{opt}}{\varphi} \quad (7)$$

The viscosity of the ferrocolloid under the influence of magnetic field of known flux density can be theoretically predicted from the above discussion. However, the predictability is limited to a particular flux density due to the inherent nature of the ferrocolloid to attain magnetic saturation. Beyond saturation, the structural rigidity of the chains seizes to enhance as a function of increasing field intensity, and as such the viscosity attains a saturation magnitude. The accuracy of predictability provided by the present mathematical approach has been illustrated in Fig. 4(b) with respect to present experimental observations. The predictability for Nickel oxide–Ferrous Oxide nanocomposite based oleo-ferrofluids[12] has also been illustrated in Fig 4(c).

**FIGURE 4**

## 3.3. Field Induced Structure Formation: Manifestation as Yield Stress

The structure formation of the SPION colloid in presence of magnetic field also manifests itself though the increasing magnitude of the yield stress the colloid can sustain. An analytical model to mathematically predict the field induced yield stress from magnetic and material properties of the ferrocolloid and its constituents has also been proposed in the present article. Experimental observations[21] suggest that the field induced augmented yield stress of magneto rheological fluids and ferrocolloids exhibit a quadratic relationship with the applied magnetic field strength. Most models existent in literature involve chain aggregate formation governing non–linear differential equations[32] that need to be solved by extensive numerical simulations. However, analytically derived models to predict the yield stress of a particular ferrocolloid under the influence of magnetic field of known strength is as of non-existent in scientific reports. The yield stress in such systems is an important magneto–mechanical property since the applicability of such colloids is decided by its load bearing capacity; a direct manifestation of the yield stress. In the present approach, an assumption (verified from reported experimental findings) that the yield stress is a quadratic function of the field strength has been made. The form of the equation involving the yield stress ($\sigma_y$) and the magnetic field strength ($B$) is expressed as

$$\sigma_y(B) = c_1 B^2 + c_2 B + c_3 \tag{8}$$

Since the ferrocolloid exhibits Bingham plastic characteristics even in the absence of external magnetic fields, mathematical analysis predicts that the magnitude of the constant '$c_3$' should be the value of the yield stress of the particular ferrocolloid of interest under zero field ($\sigma_{y,zf}$). The final form of Eqn. (1) thereby becomes

$$\sigma_y(B) = c_1 B^2 + c_2 B + \sigma_{y,zf} \qquad (9)$$

The factors that hold appreciable promise as potential governing parameters of the field augmented yield stress are the nanoparticle magnetic moment at a given field strength, the nanoparticle dimensions, the characteristic dimensions of the field induced magnetic chain structures formed within the ferrocolloid, the magnetic permeability of the base fluid and the nanoparticulate material, the force experienced by the nanoparticles within the magnetic field while forming aligned chained structures, etc. The analytical model encompasses all the contributing factors and expresses the magnitudes of the coefficients in terms of the above factors.

The coefficient '$c_2$' relates the applied magnetic field to the induced yield stress. Augmentation in yield stress involves formation of steady chain structures composed of a finite number of magnetic particles per chain. It can be argued from intuition that the yield stress magnitude of the ferrocolloid is directly linked to the magnitude of the magnetization per chain ($M_c$) for a given field strength. The validity of the assumption can be derived from the argument that the total magnetic moment of a singular chain structure provides a direct insight into the shear resistance the colloid can provide in presence of a filed. As stated in the previous paragraphs, it can be theorized that the total strength offered by the colloid in presence of magnetic field is inversely related to the average length of the chains formed and the particle separation distance. As observed experimentally and reported[32], the yield stress is related directly to the square root of the concentration ($\varphi$). Based on the preceding discussions, the dimensionally consistent form of coefficient '$c_2$' is expressed as

$$c_2 = \frac{M_c \varphi^{1/2}}{L_c^2 d_{c,sep}} \qquad (10)$$

The inter-particle separation can be deduced accordingly from geometrical considerations; however, it is also a function of the concentration of the magnetic particles, which can be deduced from Eqn. (7). With functional magnitudes deduced from the preceding formulations, it is now possible to determine the magnetic moment per chain structure. The average number of magnetic particles building up each chain structure can be determined from the average particle diameter and the corresponding separation at a particular concentration, which in turn provides a magnitude for the average mass of a particular chain. The product of the average mass to the magnetic moment of the sample at particular field strength provides the magnitude for the total magnetic moment available per chain under the influence of particular field strength. The average moment per chain can be expressed in terms of magnetic moment of the sample ($M_s$), particle diameter, particle density, separation distance and average chain length as

$$M_c = \frac{M_s \rho_p d_p^3}{6/\pi} \left( \frac{L_c}{d_p + d_{c,sep}} \right) \qquad (11)$$

The mathematical expression for the coefficient '$c_1$' can also be deduced on similar grounds. In order to obtain a dimensionally consistent form of the coefficient, it can be mathematically deduced that it is essentially a force per unit area per unit magnetic field strength squared. Since the force term comes in existence during the existence of the magnetic field, it can therefore be considered to be the inter-particle magnetic force that leads to the formation and stability of the chained structures. For singular domain magnetic nanoparticles, it has been verified[4] that the magnetic force between two particles follows the inverse square law similar to Coulomb's law for point charges. Just as the electrostatic forces between two objects increase with decreasing distance due to divergence of the electric field within the bodies, the magnetic field in between two particles increase tremendously up to the saturation magnetization upon chain formation. Therefore, for concentrated systems, wherein the interparticle separation can be expected to be of the order of particle diameter, the saturation magnetization can be utilized to deduce the interparticle magnetic force. The magnitude of the force ($F_m$) is expressed as[4]

$$F_m = 3\mu_f d_p^2 \beta^2 H^2 f \qquad (12)$$

where, '$\mu_f$' represents the permeability of the fluid, '$H$' the magnetic field strength, '$\beta$' the equivalent of the electrostatic Claussius–Mossotti factor for magnetic dipoles and '$f$' the particle separation function. The factor '$\beta$' can be expressed in terms of '$\alpha$', the ratio of permeability of the particle material to that of the fluid as

$$\beta = \frac{\alpha - 1}{\alpha + 2} \qquad (13)$$

From the mathematical expression for '$f$', it can be shown that for a system as '$d_{sep}/d_p$' tends towards zero; the value of the factor reduces to twice the fourth power of the ratio of the particle diameter to that of the inter-particle separation. In the present case, since the nanoparticles are super paramagnetic, they overlap in presence of external fields to form strong chains, thereby leading to the separation distance being of the order of particle diameter, or lower, i.e. leading to values of the ratio tending from unity towards zero. The final expression for '$f$' is as follows

$$f = 2\left(\frac{d_p}{d_{c,sep}}\right)^4 \tag{14}$$

where, the magnitude of the ratio of SPION diameter to separation can be deduced from Eqn. (7).

Although the effects of concentration are aggregated within the formulation for the permeability of the ferrocolloid, its effective magnitude of '$c_1$' needs to be intuitively modeled. As the concentration of particles increases, the propensity of the particles to experience forces from the adjacent chains grow in magnitude due to the increased population density of particles. As a result, anisotropy within the polar force field increases and reduces the absolute chain strength. Although the shear resistance increases due to increased chain population, the ultimate shear resistance capacity decreases due to weakening of the chains. This can be supported by the fact that beyond a certain optimal concentration, the magnetic response of the colloid decreases and phase separation begins even at low fields. Thereby, a simple inverse relationship for the polar magnetic force function with the concentration has been established and has been found to be consistent for all experimental observations. Converting the magnetic field strength to the available magnetic flux density within the ferrocolloid ($B_c$) and reducing to the dimensional form required to obtain a dimensionally consistent equation by combining Eqns. (12), (13) and (14), the final expression for the force between the magnetic particles within the magnetic field is as follows

$$c_1 = \frac{6\mu_f}{\varphi \mu_c^2}\left(\frac{d_p}{d_{c,sep}}\right)^4 \left(\frac{\alpha-1}{\alpha+2}\right)^2 \tag{10}$$

The term '$\mu_c$' denotes the permeability of the ferrocolloid. Although the system is supposed to exhibit highly non-linear behavior, a linear model can be utilized to obtain the effective permeability of the colloid, justified by the fact that the colloids are restricted towards the lower side of concentration and are highly viscous to dampen out nonlinear effects caused by extensive particle migration. The permeability can be modeled based on effective property approach from the permeability of the particle material '$\mu_p$' and that of the base fluid '$\mu_f$' as

$$\mu_c = \varphi \mu_p + (1-\varphi)\mu_f \qquad (11)$$

However, for low concentration systems, the Maxwell–Garnett model (Eqn. 12) may be utilized to obtain effective permeability values with higher degree of accuracy.

$$\frac{\mu_c - \mu_f}{\mu_c + 2\mu_f} = \varphi \frac{\mu_p - \mu_f}{\mu_p + 2\mu_f} \qquad (12)$$

The ferrocolloid as well as the base fluid behaves as non-ideal Bingham plastics and exhibit yield stresses under shear. However, analysis reveals that the high shear deformation is not purely linear but exhibits certain amounts of minimal non–linear behavior. It can be therefore argued that the present ferrocolloids are practically Hershel–Buckley fluids with very low values of the viscosity index parameter and the viscosity consistency parameter having magnitude equal to the real viscosity of the system. The shear deformation characteristics and increase in yield stress as a function of magnetic field strength and concentration and the validity of the proposed model have been illustrated in Fig. 5 (a) and (b) respectively.

**FIGURE 5**

Figure 5(b) has also been plotted in a log-log format (not illustrated) and the changes in slope were observed. It has been reported[33] that for MR fluids, the slope remains ~ 2 for low magnetic fields while transits to ~ 1.5 at higher fields. However, for the present colloids, the slope was found to remain nearly constant at 2 even at higher magnetic fields. This might be caused by the fact that the particles are nanoscale and therefore less prone to be affected by magnetic saturation at high fields compared to micron scale particles in conventional MR fluids. It is possible that the effects of saturation start manifesting itself at much higher fields when compared to the critical field for conventional MR fluids.

### 3.4. Reversible magneto-viscosity

The reversibility of magneto-viscosity is one of the more prominent features desired of such ferrocolloids for application point of view. A stable and high response colloid should possess the ability to increase its viscosity and return back to the initial state under the influence of an external field in the shortest possible time period and with least viscous hysteresis. In the present scenario, viscosity has been recorded as a function of increasing and decreasing magnetic field with varying time periods of actuation. The transient response of the ferrocolloid as a function of increasing and decreasing magnetic field cycle with a cycle time of 0.5 second has been illustrated in Fig. 6. At this cycle time, the ferrocolloid exhibits negligible hysteresis and the initial and final viscosities converge perfectly. The hysteresis observed at high magnetic flux densities is a consequence of small time period available in order to revert back to the initial state from the field induced microstructure of the ferrocolloids. However, at low fields, the induced structural rigidity is much lower and hence the hysteresis completely disappears. The synthesized SPIONs are thus in a position to orient the whole colloidal system along the magnetic field within a very short time frame and render the colloid high degree of reversibility.

**FIGURE 6**

## 4. Field Induced Viscoelastic response of the ferrocolloid

Oscillatory tests have been performed to determine the viscoelastic behavior of the ferrocolloid in presence of external field stimulus. The viscoelastic behavior adds further insight into the field induced structure formation and its implications. Oscillatory shear experiments at constant frequency (10 Hz) have been carried out to determine the linear viscoelasticity (LVE) region in the strain amplitude range of 0 to 20%. The linear regimes; where storage ($G´$) and loss ($G´´$) moduli are constant regardless of strain amplitude have been clearly observed. Similar viscoelastic response has been observed for the base PEG–nSi complex with the LVE extending up to 0.15 % strain amplitudes (not illustrated). Also, $G´´$ is observed to be greater in magnitude than $G´$ for the base complex and the ferrocolloid (in absence of field) and therefore they essentially exhibit higher fluid characteristics than solid characteristics; i.e. they are viscoplastic[34] rather than desired viscoelastic and therefore cannot resist shear deformation. The objective lies in increasing the elastic nature of the colloid by applying external magnetic stimuli, which is often lacking in many literatures in the field. Experiments have been performed at various magnetic flux densities to examine the effects of field induced structure formation on the viscoelastic properties of the ferrocolloid. The storage and loss moduli curves for 10 wt. % SPION ferrocolloids under variant magnetic fields have been illustrated in Fig. 7 (a) and (b) respectively.

As illustrated in Fig. 7, the ferrocolloid exhibits enhanced field tunable viscoelastic moduli; especially G´. It may be noted that the base fluid and the ferrocolloid exhibits LVE in a very narrow strain amplitude belt reaching up to a maximum of 0.25 %. It has been observed that the LVE region shifts towards higher strain amplitudes increasing external field strength, as has been illustrated by a dotted line in Fig 7 (a) and (b). This implies that the ferrocolloid experiences field induced structure formation which is strong enough to withstand the oscillatory shear to a greater extent than that at lower field intensity. The LVE region indicates zero or minimal microstructural deformation caused by shear loading. The microstructural deformation in this region, if any, is so minimal that the colloid can return back to its initial state in equilibrium with the imposed shear, thereby exhibiting no change in the magnitude of its viscoelastic moduli. The moduli then undergo monotonous decrease in magnitude with increase in strain amplitude, i.e. no stable response is further obtained. This drop in magnitude after a critical strain is due to inability of the microstructure to regain its structural state in equilibrium with the imposed shear; thereby leading of mechanical instability. The present ferrocolloid thereby lodges field induced microstructure strong enough to extend the elastic region considerable by ~16 times at a SPION concentration of 10 wt. % under a magnetic field of 1 T. The storage modulus obtained is also very high for SPION based colloids, which further speaks of the high stability of the ferrocolloid caused by the presence of the PEG–nSi complex.

**FIGURE 7**

Another important aspect of the present ferrocolloid lies in its field induced augmentation of the crossover strain amplitude. This happens to be another important property which is often neglected in many reports in the field. The crossover amplitude indicates the strain at which the colloid ceases to retain its solid–like character and transits to a more liquid form. The present work illustrates the impact of magnetic field and SPION concentration on the crossover amplitude in Fig. 8.

**FIGURE 8**

The enhanced crossover amplitude with increasing concentration speaks of increased solid–like characteristics retention with increase in SPIONs. The elastic nature imparted by the presence of granulated media in the base matrix leads to enhanced solid character retention. The crossover amplitude is also a function of the externally applied magnetic field intensity, thereby confirming field induced structure formation. However, the behavior is similar to the field induced augmentation in viscosity and attains a plateau value at certain strength of external field. The phenomenon of plateauing is caused by the limitation of the field induced microstructure to enhance the structural population further. As discussed earlier, the SPION population within the colloid arrange themselves into chain structures in presence of a magnetic field in way such that thermally stable chains, with the maximum possible chain population density are formed. In other words, the thermal and magnetic field leads to the formation of microstructure optimized for maximum allowable shear resistance. As a result, at a given concentration of SPIONs, the ferrocolloid, governed by the saturation magnetization of the SPION material and the chain formation optimization can attain a maximum solid–like character. Beyond this point, the shear resistance of the colloid fails due to its inability to attain further structural strength. This restricts the maximum attainable crossover strain amplitude to a concentration and field dependent plateau magnitude, '$\varepsilon_{c,p}$'. Analysis shows that the crossover strain amplitude '$\varepsilon_c$', in percentage holds a decaying logarithmic relationship with the external field intensity '$H$' as

$$\varepsilon_c - \varepsilon_{c,p} = A \ln H \qquad (13)$$

In Eqn. 13, the magnitude of the variable '$A$' can be determined empirically from analyses of data and has been observed to be equal to the SPION concentration (in wt. %). The logarithmic plateauing is expected as any charge growth phenomena since the propensity of formation of chains of a particular length and population density are directly dependent on the concentration and field intensity as well as the saturation magnetization of the SPION material. It is similar to any charge growth system; wherein the propensity of charge accumulation reduces with increasing charge content due to the decreasing potential between the system and the source of charge. In the ferrocolloidal system, the similar 'potential' is provided by the SPION material's concentration and saturation magnetization. At a given concentration, as the field intensity approaches the saturation field, the propensity of chain formation decays down due to decrease in 'magnetoviscous' potential of the colloid. Similarly, as the concentration increases at a given field intensity, the propensity to form stable chains decays due to particle overcrowding and decreasing colloid stability; i.e. decreasing 'spatial' potential.

On similar grounds, the variance of $G'$ and $G''$ of the ferrocolloid with SPION concentration and external magnetic stimuli has been experimentally determined (within the corresponding LVE) and has been illustrated in Fig. 9 (a) and (b) respectively. The observed trends are consistent with the discussion in the preceding section. Viscoelastic moduli, the quantifying parameters of elastic and viscous nature of the colloid are expected to follow similar trends as the augmentation in crossover amplitude due to similar governing mechanisms. Analysis shows that similar to the crossover amplitude, $G'$ and $G''$ also attain field and concentration governed plateau magnitudes, $G_p'$ and $G_p''$ respectively and possess a logarithmic decaying relationship with the field intensity as

$$G' - G_p' = B \ln H \qquad (14)$$

In similitude to the variant '$A$', the variable '$B$' can also be determined empirically as $\sim 10^3$ times the square root of SPION concentration (in wt. %). As explained in the previous section, the logarithmic rate of decay is an outcome of the decreasing 'magnetoviscous' potential with increasing field intensity. However, with increase in elastic nature, the colloid also gains an inherent augmentation of its ability to dissipate the stored strain energy caused by external load and as a result, the loss modulus also increases in a similar fashion, but simultaneously lower in magnitude than the storage modulus. This phenomenon reveals that the ferrocolloid augments its solid–like characteristics while keeping its dissipative potential constant. This makes it a potential candidate in short cycle vibration control applications wherein high elastic as well as quick dissipative characters (with removal of field) are desired.

**FIGURE 9**

**FIGURE 10**

For viscoelastic media, the effect of frequency of the oscillatory loading within the LVE should be examined to properly comment on the viscoelastic response of the colloids and ascertain their applicability in dynamic and transient applications. As evident from Fig. 10 (a) and (b), where the effect of oscillatory angular frequency and magnetic field intensity on $G'$ and $G''$ (strain amplitude within the LVE) have been illustrated, the ferrocolloid exhibits frequency independent solid–like character. The ferrocolloid also exhibits increase in magnitude of the viscoelastic moduli with increasing SPION concentration (not illustrated) similar to the trends observed in Fig. 9 (a) and (b) respectively. It can be observed that the storage and loss moduli increase with increasing magnetic field, however, the solid character remains more than the fluid nature. The phenomenon is important since it sheds light onto the stability of the field induced chain structures. Increasing frequency of oscillating load indicates decreasing time period for the field induced chains to relax back to their initial spatial position and orientation, thereby indicating a probability of decrease in the field induced elastic characteristics of the ferrocolloid. However, the present data reveals that despite the decrease in time period available for relaxation, the SPION based chained structures can actively respond to the stimulus and regain their initial spatial distribution and orientation. Such behavior indicates high response of the PEG–nSi complex stabilized SPION based ferrocolloid under strong transient and or dynamic loading and warrants the present material as potential candidates for high frequency MEMS/NEMS actuators and/or sensors.

## 5. Conclusion

High stability ferrocolloids with SPIONs embedded in a PEG–nSi base matrix have been synthesized and magnetorheological studies have been performed to understand the internal mechanics of the colloidal microstructure in presence of magnetic field. The ferrocolloids reveal high stability and reversibility up to 15 wt. % of SPION loading under magnetic field, even as high as ~1 T. The effects of SPION concentration on the augmented magnetoviscosity and yield stress under magnetic field has been observed and mathematically modeled based on fundamentals of magnetic saturation, chain structure formation and strength, magnetic anisotropy and Neel–Brownian relaxation. The proposed models, analytical in nature, have been found to accurately predict the magnetoviscous phenomena for all concentrations and external field strength. The present article also includes reports on the magnetoviscoelastic and hysteresis of the ferrocolloids. The colloids have been demonstrated to possess high response to transient fields and high solid character under oscillatory loads. Field tunable enhancements of storage modulus and LVE have been observed and the phenomena have been explained based on micromechanics of magnetic particles suspended in fluid under magnetic field. The proposed explanations and mathematical analysis have been observed to accurately predict the various magnetoviscous properties from knowledge of SPION concentration, magnetic properties of the SPION material and the applied field intensity. The reported ferrocolloids exhibit properties that show good promise for direct applications in magnetic NEMS/MEMS devices, seismic damping, dynamic automotive components, oleo sealing and nanomanufacturing.

## Acknowledgment

The authors acknowledge the Defense Materials and Stores Research and Development Establishment (DMSRDE, India) (a Defense Research and Development Organization (DRDO) lab) for funding the present research initiative. The authors thank Dr. V Ramanujachari, Director, DRDO RIC, IIT Madras Research Park, Chennai and Dr. Arvind K. Saxena, Director, DMSRDE, Kanpur for the fruitful technical discussions. The authors would like to thank the staff of the Sophisticated Analytical Instruments Facility, IIT Madras for assisting in characterization of the SPIONs. P D would also like to thank the Ministry of Human Resource and Development (MHRD), Govt. of India, for the doctoral research scholarship.

**FIGURE CAPTIONS**

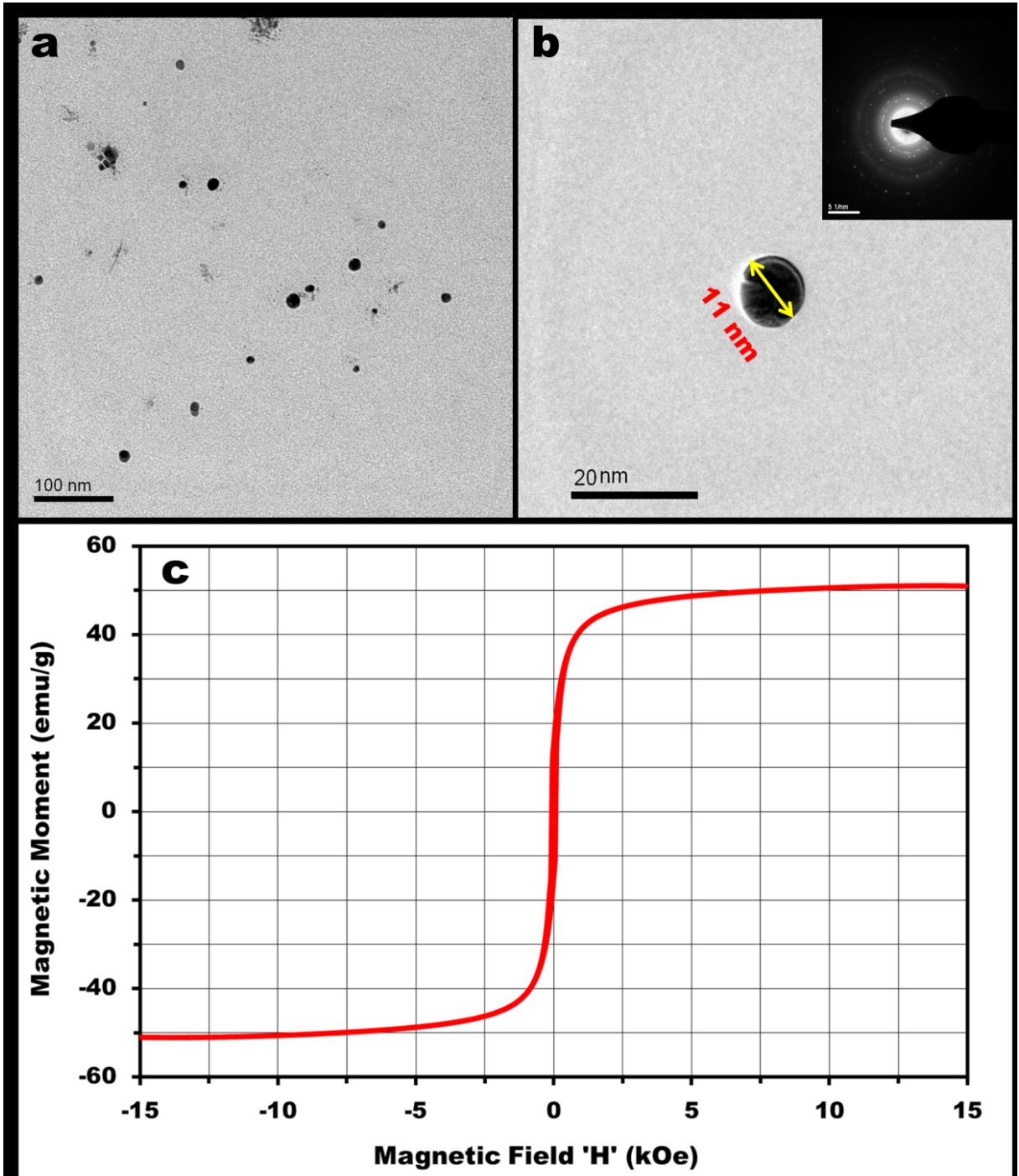

**Figure1.** **(a)** Transmission Electron Microscopy (TEM) image of synthesized SPION. The size range is observed to be 5-15 nm and thereby the particles are super paramagnetic in nature (Critical size for $Fe_3O_4 \sim 150$ nm) **(b)** TEM image of single nanoparticle for size determination **(Inset)** Selected Area Diffraction Pattern (SAED) of sample showing crystalline nature **(c)**

Vibrating Sample Magnetometer (VSM) measurement of SPION at 300K. The saturation magnetic moment is determined to be 51 emu/g and the saturation field ~ 12.5 kOe. No hysteresis loop is observed which indicates the super paramagnetic nature.

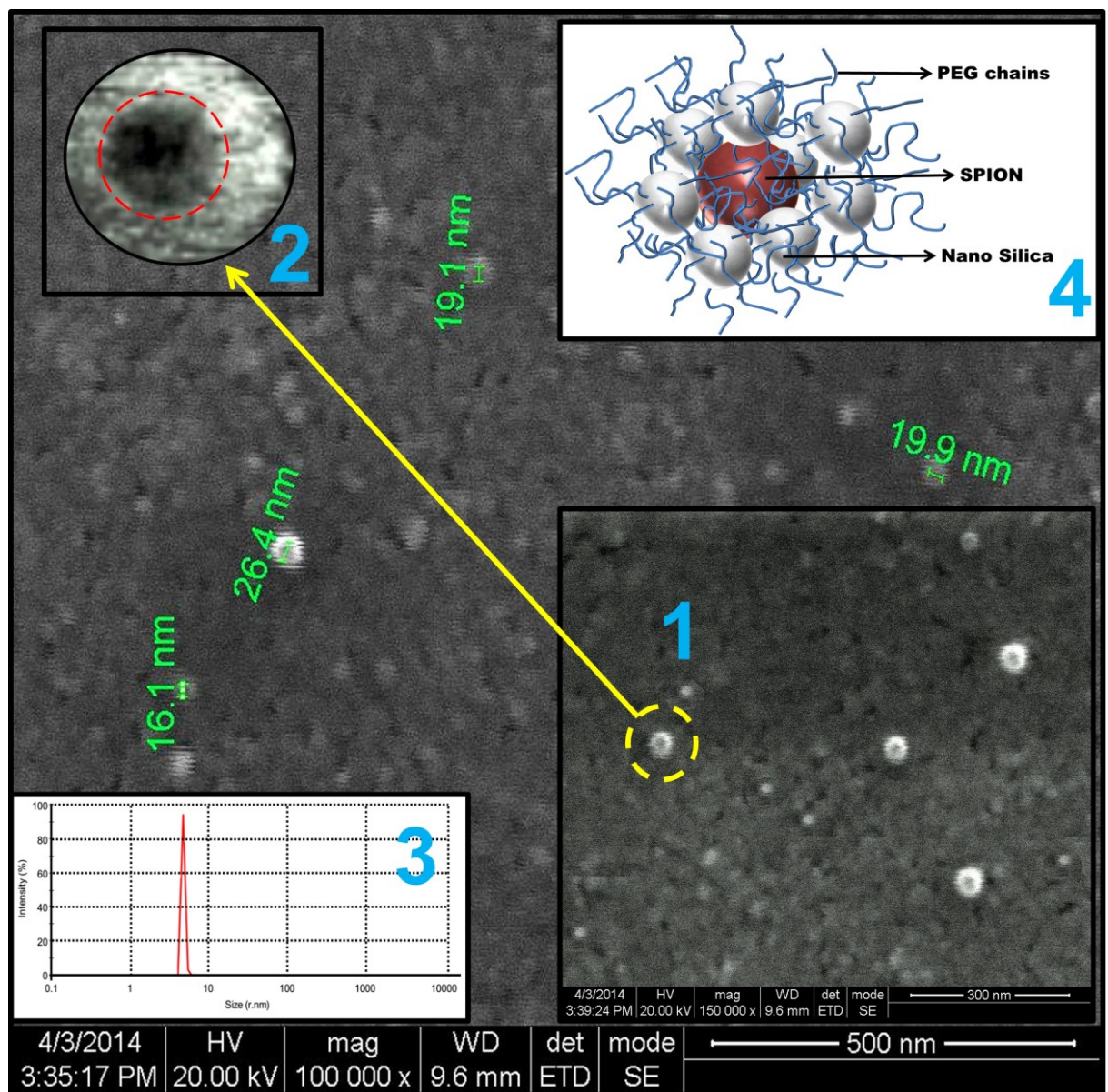

**Figure 2:** HRSEM characterization of the ferrocolloids. The ultrasonication gradually coats the SPIONs with a nanoscale layer of the PEG-nSi complex and the final diameter of the coated SPIONs range within ~15 –25 nm. **Inset 1** shows an enlarged view of the dispersed SPIONs in the dry PEG-nSi complex matrix. Clustered silica nanoparticles dispersed in the PEG are also visible as scattered but prominent white dots within the matrix. **Inset 2** shows a magnified view of a single SPION to reveal the presence of an optically rarer stabilizing coat over the denser magnetic particle. The formation of the coat provides the high stability and response to the ferrocolloids and prevents phase separation even under high magnetic fields and imposed shear. **Inset 3** shows the DLS data for the nano silica utilized. The nano-silica can also be seen

scattered within the SEM image matrix as distinguishable minute light shade pixels. **Inset 4** shows the probable mesh network structure formed by PEG-nSi which traps the SPIONs, leading to highly stable colloids even under the influence of high magnetic fields.

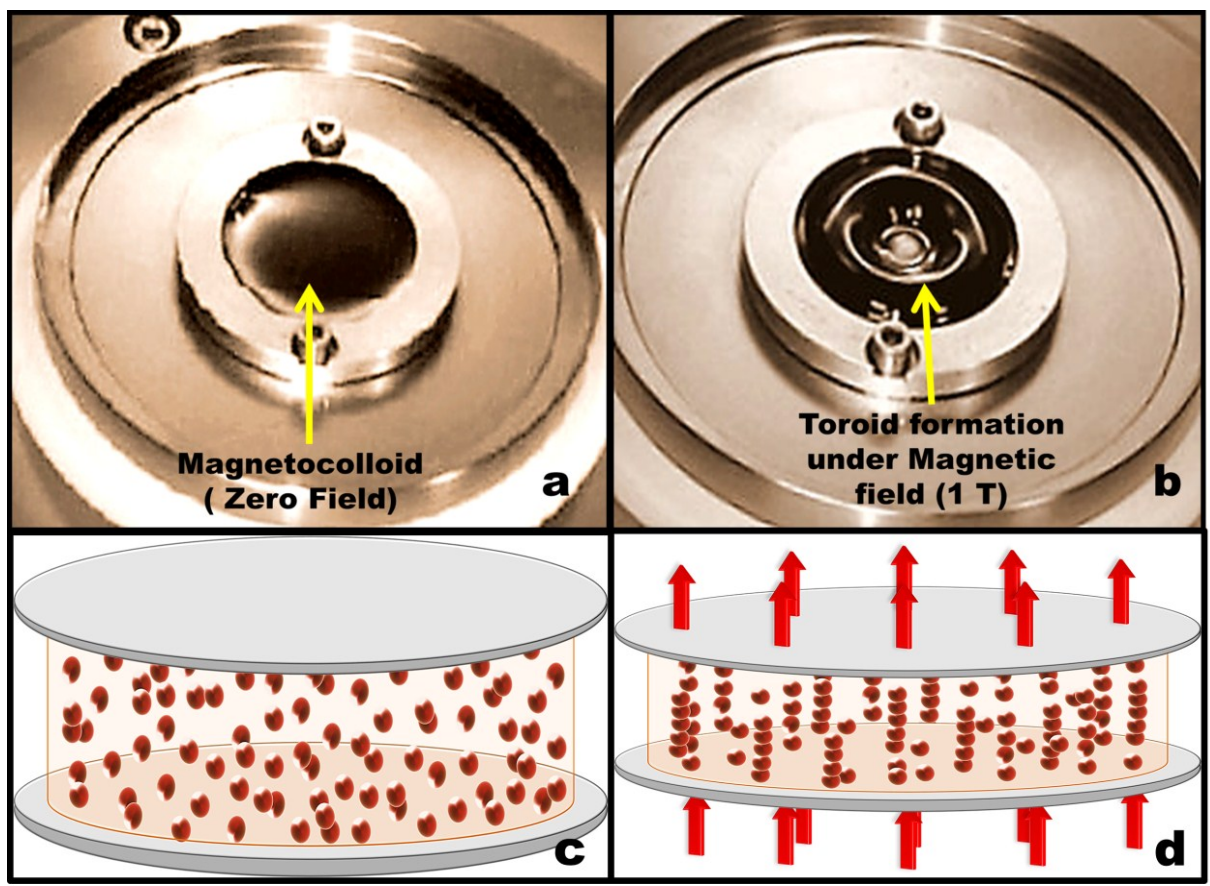

**Figure 3: (a)** PEG-nSi-SPION (10 wt. %) ferrocolloid in absence of magnetic field **(b)** increased solid character and toroidal structure formation under 1 T magnetic field. observed at high field vanishes. **(c)** and **(d)** illustrate qualitatively the mannerism in which the SPIONs form chained structures in presence of the applied field (denoted by red arrows), which in turn leads to the transition from (a) to (b). The chained structures deploy enhanced structural rigidity to the colloid, thereby increasing shear resistance.

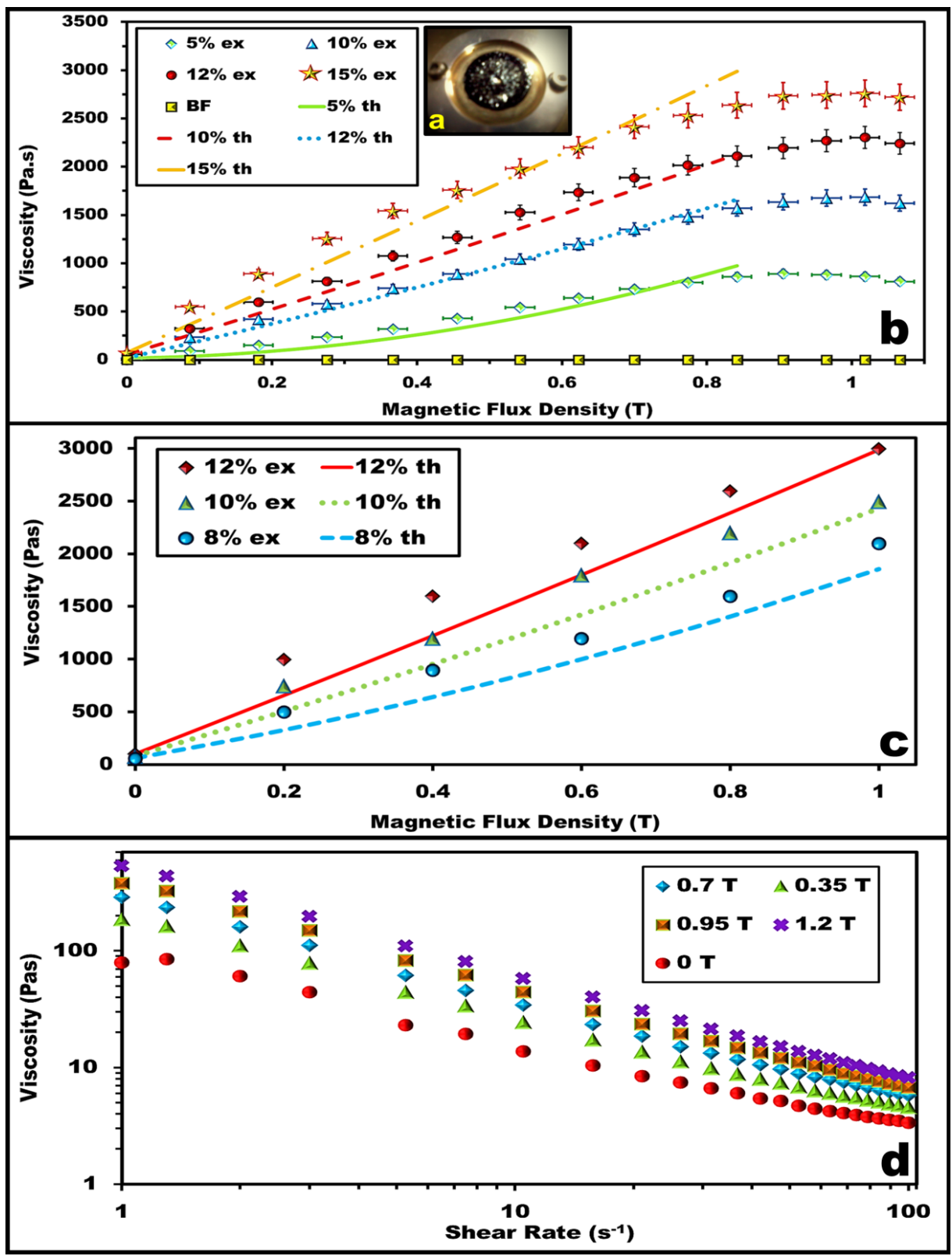

**Figure 4:** Magneto-viscous effect exhibit by the ferrocolloid as a function of the applied magnetic flux density. The illustration **(b)** provides values of the magneto–viscosity measured at a constant shear rate of 1 s$^{-1}$. At the highest SPION concentration corresponding to a field

stable ferrocolloid, the magneto–viscosity has been observed to be almost 4,000 times (~ 0.9 T) the viscosity of the carrier fluid viscosity. The optimal concentration has been deduced by observing the response of the colloid to magnetic fields. A 16 wt. %, the ferrocolloid can no longer regain back the fluid nature upon withdrawal of the field, as seen in inset **(a)**. **(c)** The present analytical model also accurately predicts the magnetic viscosity for ferrofluids (experimental results published in literature[12]). Such high magneto-viscosity coupled with low relaxation time, low hysteresis and high stability shows good promise for potential applications. **(d)** Illustration of the dependence of viscosity of 10 wt. %. SPION ferrocolloid with respect to imposed shear and magnetic field.

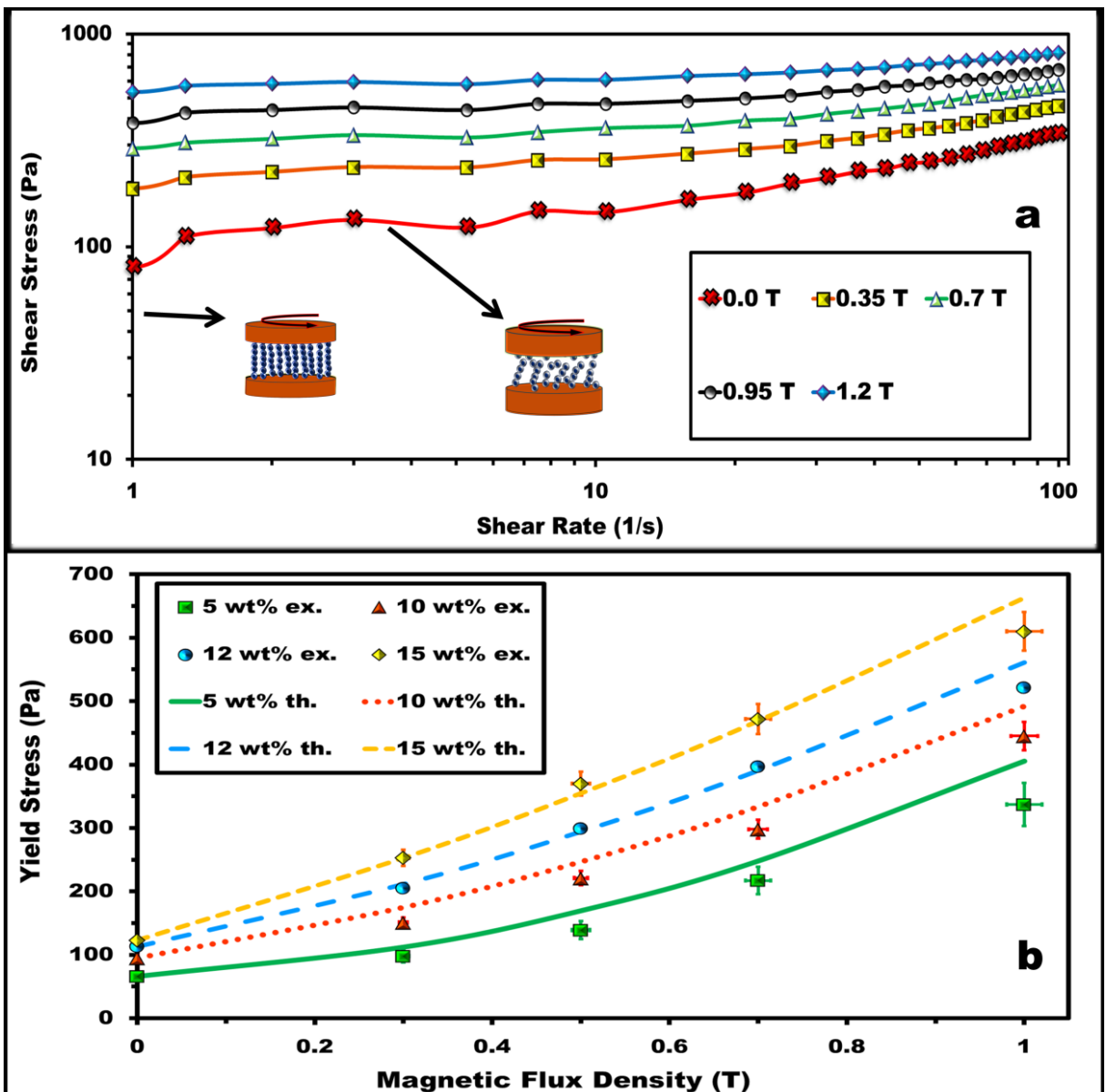

**Figure 5: (a)** Augmented yield stress of the ferrocolloid under the influence of external magnetic field due to structural rigidity caused by chain formation. The ferrocolloid behaves like

a Bingham plastic. All lines are for guide to the eyes. Illustrations directed with arrows provide a qualitative picture of the chain structures within the shear resisting and the continuous deformation regimes. **(b)** The present mathematical model predicts the yield stress value at a particular concentration and field intensity accurately, hinting at the fact that the mechanisms considered for the analysis are the most probable ones responsible for nanoscale magnetorheological response.

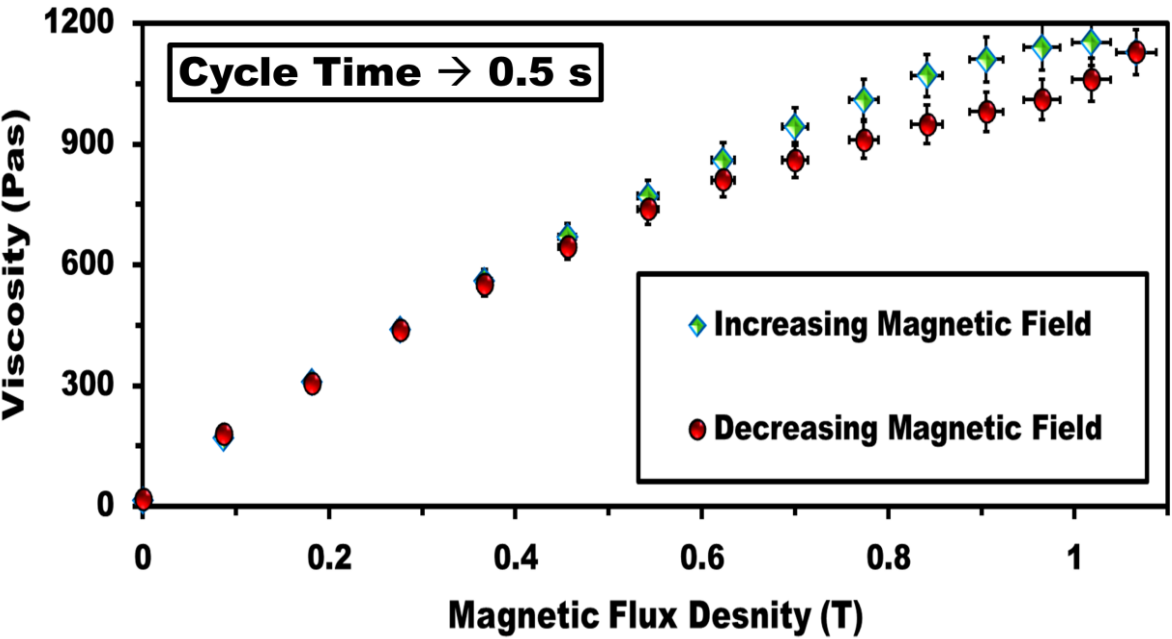

**Figure 6:** Reversible nature of the magnetoviscosity exhibit by the ferrocolloid (10 wt. % SPION) during impulsive field cycles.

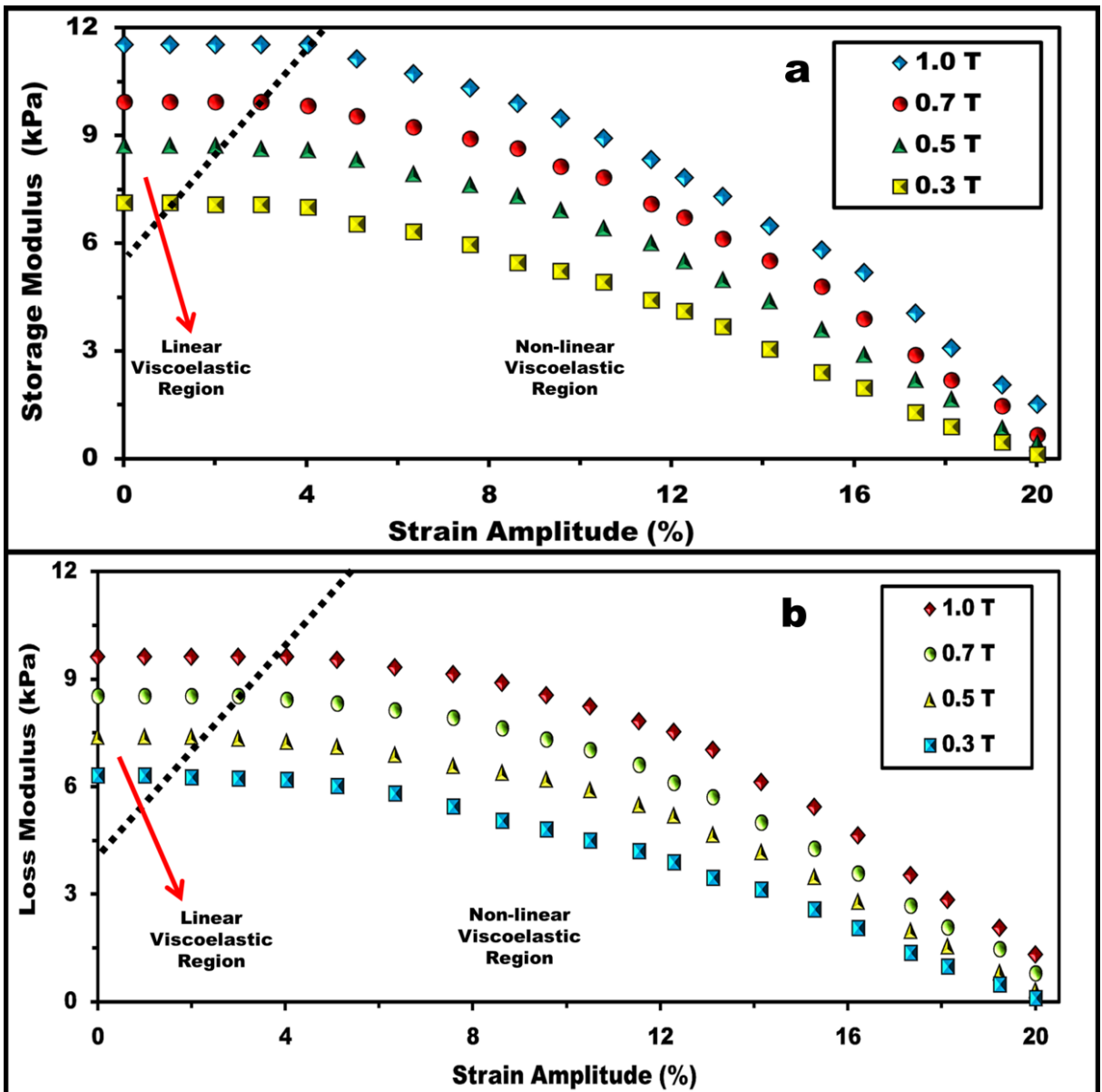

**Figure 7:** Augmentation in viscoelastic moduli; **(a)** storage and **(b)** loss modulus of the ferrocolloid (10 wt. % SPION) as a function of magnetic flux density and applied strain amplitude. The enhancement in LVE with increasing magnetic field intensity has been demarcated with the dotted line.

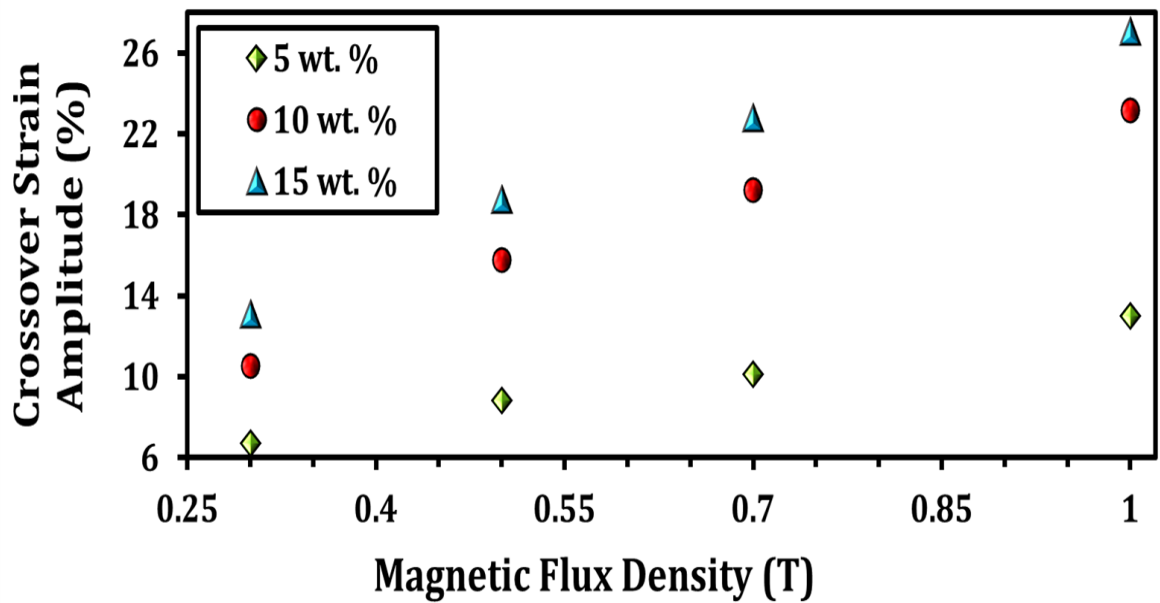

**Figure 8:** Enhancement of the crossover strain amplitude of the ferrocolloid as a function of SPION concentration and external field intensity. The colloid has been observed to exhibit crossover amplitude of ~ 26% strain at 15 % SPION loading and 1 T external field. Such behavior represents transition to sustained solid like state with increasing concentration and magnetic field.

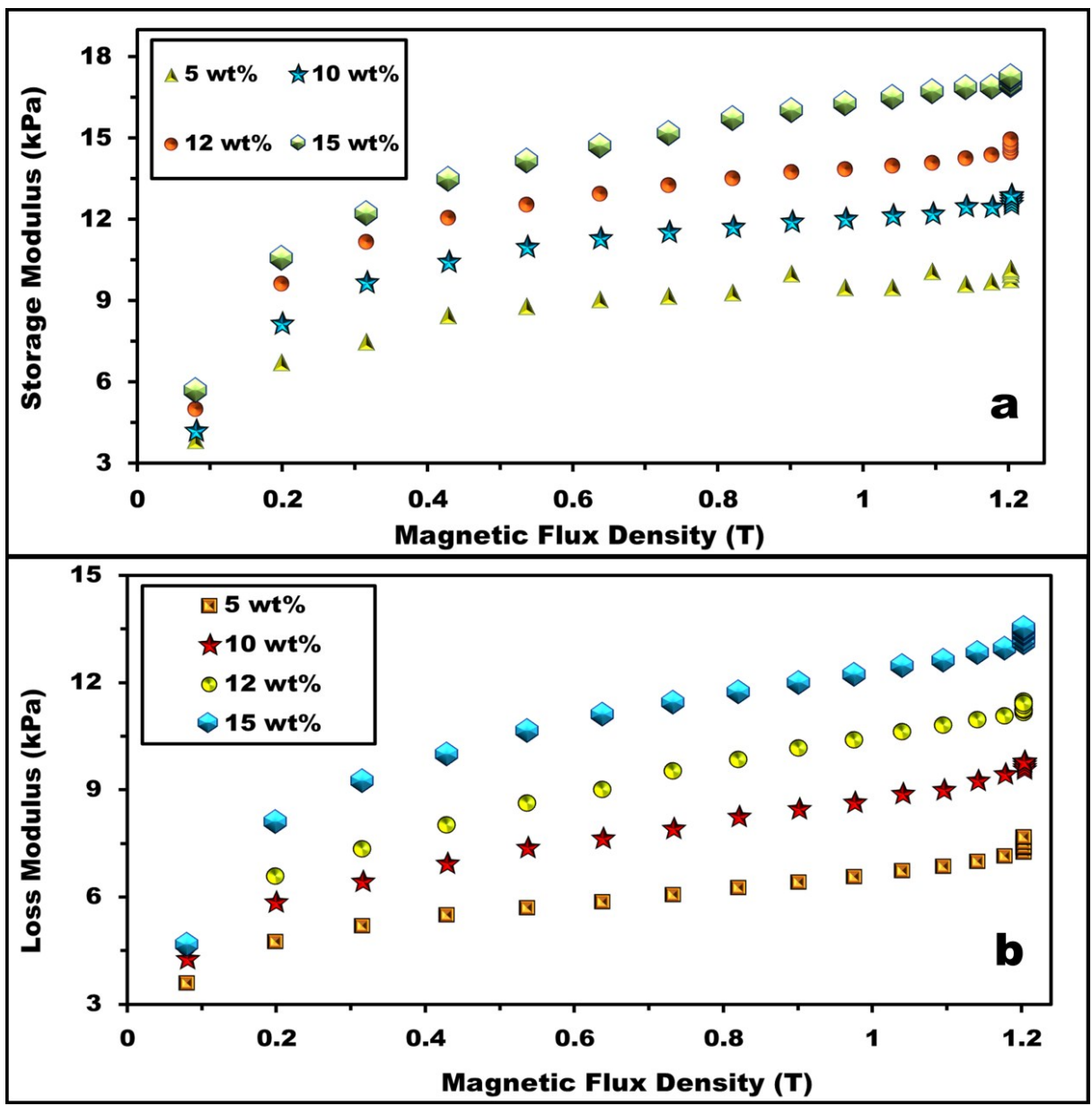

**Figure 9:** Variation of viscoelastic moduli (at constant oscillatory strain of 1% and frequency of 10 Hz); **(a)** storage and **(b)** loss modulus as a function of SPION concentration and magnetic flux density.

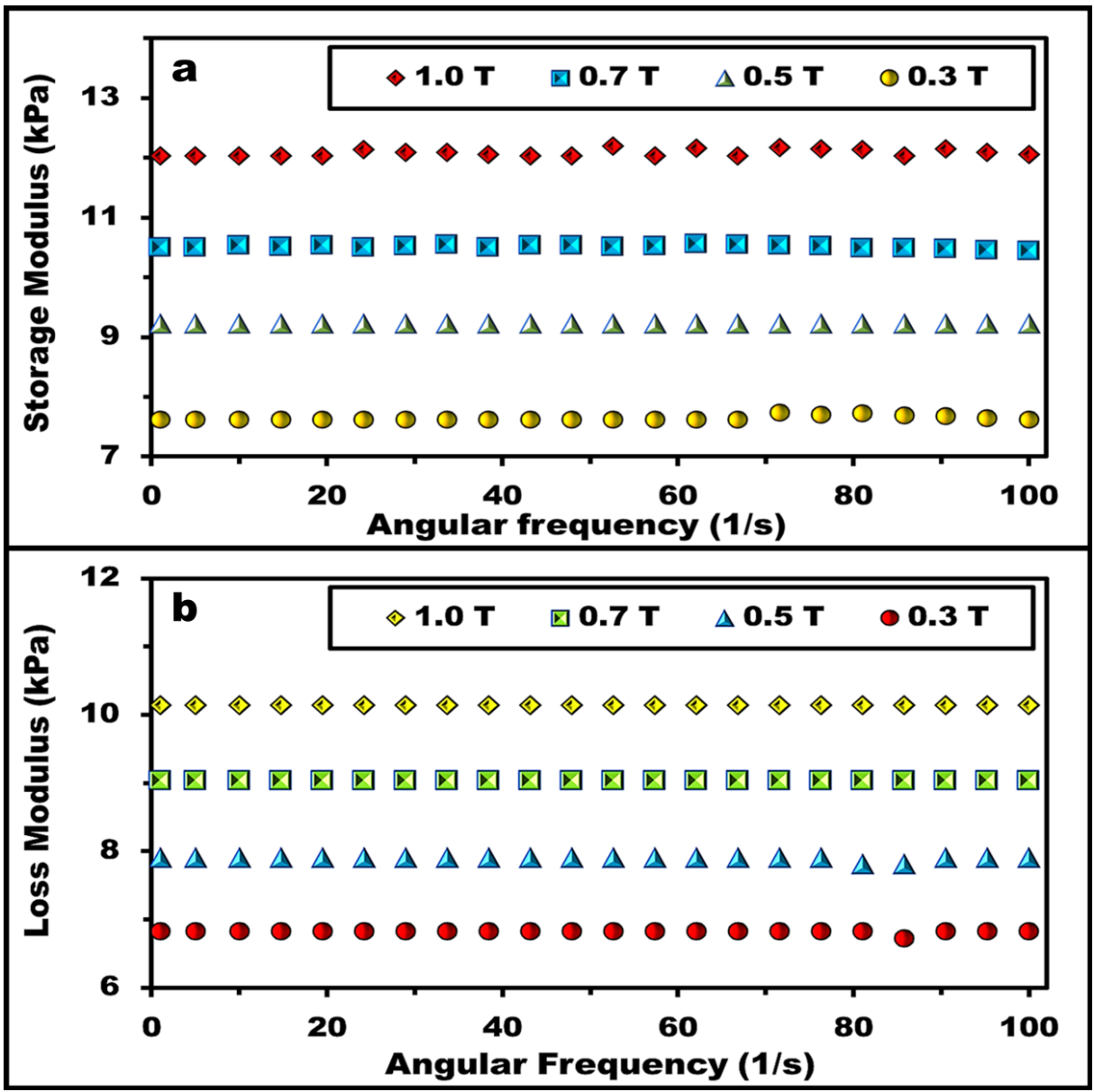

**Figure 10:** The variation of viscoelastic moduli (at an oscillatory strain of 1%, i.e. within the LVE); **(a)** storage and **(b)** loss modulus as a function of frequency of imposed oscillatory load and applied magnetic flux density. The ferrocolloid shows constant viscoelastic moduli with respect to frequency, thereby confirming that it can retain its solid like character irrespective of forcing frequency.